\documentclass{wscpaperproc}
\usepackage{latexsym}
\usepackage{graphicx}
\usepackage{mathptmx}

\usepackage{amsmath}
\usepackage{amsfonts}
\usepackage{amssymb}
\usepackage{amsbsy}
\usepackage{amsthm}

\newtheoremstyle{wsc}
{3pt}
{3pt}
{}
{}
{\bf}
{}
{.5em}
{}

\theoremstyle{wsc}

\newtheorem{example}{Example}

\begin{document}

%
%
\WSCpagesetup{Christian P. Robert}

\title{SIMULATION IN STATISTICS}

\author{Christian P. Robert\\ [12pt]
Universit{\'e} Paris-Dauphine \\
CEREMADE \\
75775 Paris cedex 16, France \\
}

\maketitle

\section*{ABSTRACT} Simulation has become a standard tool in statistics because it may be the only tool
available for analysing some classes of probabilistic models. We review in this paper simulation tools that
have been specifically derived to address statistical challenges and, in particular, recent advances in the
areas of adaptive Markov chain Monte Carlo (MCMC) algorithms, and approximate Bayesian calculation (ABC) algorithms.

\noindent
{\bf Keywords:} Monte Carlo, MCMC, ABC, SMC, importance sampling, bootstrap, maximum likelihood, Bayesian
statistics, posterior distributions, Bayes factor, model choice, adaptivity, Markov chains.

\section{INTRODUCTION}
\label{sec:intro}

Why are statistics and simulation methods so naturally intertwined?  Statistics being based upon probabilistic
structures, stochastic simulation appears as a natural tool to handle complex issues related to those
structures, besides appealing to the statistician's intuition.  The emergence of the subfields of computational
statistics (\citeNP{gentle:2009}, \citeNP{gentle:hardle:mori:2011}) is strongly related to the rise in the use of
Monte Carlo methods and the history of statistics in the past 30 years exhibited a series of major advances
deeply connected with simulation, whose description is the purpose of this tutorial.  The coverage is
necessarily limited and the interested reader is referred to the above references and to
\citeN{robert:casella:2004} for books on simulation in statistics.


\section{MONTE CARLO METHODS IN STATISTICS}\label{sec:MCinS}

\subsection{History}

Simulation has appeared at the very early stages of the development of statistics as a field. Even though it is
difficult to find a connection between Pierre-Simon de Laplace or Carl Gau\ss~and simulation or between Charles
Babbage or Ada Byron and statistics, the great Francis Galton devised mechanical devices to compute estimators
and distributions by means of simulation. Not only his well-known quincunx (as discussed in
\citeNP{stigler:1986}) is a derivation of the CLT for Bernoulli experiments, but 
\citeN{stigler:2010} shows that Galton also found a way to simulate from posterior distributions by an ingenious
3D construct. Closer to us, the randomised experiments of Ronald Fisher \cite{fisher:1935} and the bootstrap
revolution started by Brad Efron (covered below) are intrinsically connected with calculator and computer
simulations, respectively. While Monte Carlo methods extend much further than the field of Statistics and while
giants outside our field greatly contributed to those methods (see, e.g., \citeNP{halton:1970}), some specific
computational methods have been developed by and for statisticians, bootstrap being the ultimate illustration.

\subsection{Statistical Methods}

Since the primary goal of statistics is to handle data, this field is difficult to envision without the use
of numerical methods at one level or another. The growing complexification of data, models, and techniques,
means that those numerical methods are becoming more and more central to the field, and that simulation methods
are in particular more and more a requirement for standard statistical analyses. Here are a few examples,
further illustrations being provided by \citeN{robert:casella:2004} and \citeN{gentle:2009}.

The bootstrap was introduced by \citeN{efron:1979} as a way of conducting inference for complex
distributions or for complex procedures without having to derive the distribution of interest in closed form.
Its appeal for this tutorial is that it simply cannot be implemented computer simulation. The
idea at the core of bootstrap is that, given a sample $(x_1,\ldots,x_n)$, the empirical cdf
$$
\widehat F(x) = \sum_{i=1}^n \frac{1}{n} \mathbb{I}_{x_i\le x}
$$
(where $\mathbb{I}_A$ is the indicator function for the event $A$) is a converging approximation of the true
cdf of the sample. Therefore, if the distribution of a procedure $\delta(x_1,\ldots,x_n)$ is of interest,
an approximation to this distribution can be obtained by assuming $(x_1,\ldots,x_n)$ is an iid sample from
$\widehat F$. Since the support of $\widehat F^n$ is growing as $n^n$, exact computations are impossible and
simulation is required. The simulation requirement is very low since producing a bootstrap sample
$(x^*_1,\ldots,x^*_n)$  means drawing $n$ times with replacement from the original sample. Creating a sample
of $\delta(x^*_1,\ldots,x^*_n)$'s is thus straightforward and the method can be used in an introductory course
to statistics. (The theory behind is much harder, see \citeNP{hall:1992}.)

Maximum likelihood estimation is the default estimation method in parametric settings, i.e.~in cases when the
data is assumed to be generated from a parameterised family of distributions, represented by a density function
$f(x_1,\ldots,x_n|\theta)$, where $\theta$ is the parameter. It consists in
selecting the value of the parameter that maximises the function of $\theta$, $f(x_1,\ldots,x_n|\theta)$, 
then called the {\em likelihood} (to distinguish it from the density, a function of $(x_1,\ldots,x_n)$.) Except
for the most regular cases, the derivation of the maximum likelihood estimator is a troublesome process, either
because the function is not regular enough or because it is not available in closed form. An illustration of
the former is a mixture model \cite{fruhwirth:2006,mengersen:robert:titterington:2011}
$$
f(x_1,\ldots,x_n|\theta) = \prod_{i=1}^n \{p g(x|\mu_1) + (1-p) g(x|\mu_2)\}\,,
\quad
\theta=(p,\mu_1,\mu_2)\,,
$$
where $g(\cdot|\mu)$ defines a family of probability densities and $0\le p\le 1$. This structure is usually
multimodal and numerical algorithms have trouble handling the multimodality.
An illustration of the later is a
stochastic volatility model ($t=1,\ldots,T$)
$$
x_t = e^{z_t} \varepsilon_t\,,\
z_t=\mu+\varrho z_{t-1} + \sigma\eta_t\,,\ 
\varepsilon_t,\eta_t\sim\mathcal{N}(0,1)\,,\ 
\theta=(\mu,\varrho,\sigma)\,,
$$
where only the $x_t$'s are observed. In this case, the likelihood
$$
f(x_1,\ldots,x_T|\theta) \propto \int_{\mathbb{R}^{T+1}} \prod_{t=1}^T
\sigma^{-T-1}\,\exp\left\{ -(z_t-\mu-\varrho z_{t-1})^2/2\sigma^2 \right\}\,
\exp\left\{ -x_t^2 e^{-2z_t}/2 -z_t \right\}\,\exp\left\{ -z_0^2/2\sigma^2\right\} \,\text{d}\mathbf{z}
$$
is clearly unavailable in closed form. Furthermore, due to the large dimension of the missing vector
$\mathbf{z}$, (non-Monte Carlo) numerical integration is impossible. While this specific model has emerged from
the analysis of financial data,
it is a special
occurrence of the family of hidden Markov models where similar computational difficulties strike
\cite{cappe:moulines:ryden:2004}.

Bayesian statistics are based on the same assumptions as the above likelihood approach, the difference being
that the parameter $\theta$ is turned into an unobserved random variable for processing reasons, since it then
enjoys a full probability distribution called the {\em posterior distribution.} (See \citeNP{berger:1985},
\citeNP{bernardo:smith:1994}, or \citeNP{robert:2001} for a deeper motivation for this approach, called Bayesian
because it is based on the Bayes theorem.) Inference on the parameter  $\theta$ depending on this posterior
distribution, $\pi(\theta|x_1,\ldots,x_n)$, it is then even more natural to resort to simulation for producing
estimates like posterior expectations, and even more for assessing the precision of the corresponding
estimators. For instance, the computation of the Bayes factor used for Bayesian model comparison,
$$
B_{12}(x_1,\ldots,x_n) = \dfrac{\int f_1(x_1,\ldots,x_n|\theta_1)\,\pi(\theta_1)\,\text{d}\theta_1}
{\int f_2(x_1,\ldots,x_n|\theta_1)\,\pi(\theta_2)\,\text{d}\theta_2}\,,
$$
where $f_1(x_1,\ldots,x_n|\theta_1)\,\pi(\theta_1)$ and $f_2(x_1,\ldots,x_n|\theta_1)\,\pi(\theta_1)$
correspond to two families of distributions to be compared, is most often impossible without a simulation step
\cite{chen:shao:ibrahim:2000}.

\begin{example}\label{ex:glm} In a generalised linear model \cite{mccullagh:nelder:1989}, a conditional distribution of
$y\in\mathbb{R}$ given $x\in\mathbb{R}^p$ is defined via a density from
an exponential family\index{exponential family}\index{model!generalised linear}
$$
y|x\sim\exp\left\{ y\cdot\theta(x) - \psi(\theta(x)) \right\}
$$
whose natural parameter $\theta(x)$ depends on the conditioning variable $x$,
$$
\theta(x) = g(\beta^\text{T} x)\,,\qquad\beta\in\mathbb{R}^p
$$
that is, linearly modulo the transform $g$. Obviously, in practical applications
like econometrics, $p$ can be quite large. Inference on $\beta$ (which is the true
parameter of the model) proceeds through the
posterior distribution (where\index{posterior distribution}
$\mathbf{x}=(x_1,\ldots,x_T)$ and $\mathbf{y}=(y_1,\ldots,y_T)$)
\begin{align}
\pi(\beta|\mathbf{x},\mathbf{y}) &\propto
\prod_{t=1}^T \exp\left\{ y_t\cdot\theta(x_t) 
- \psi(\theta(x_t)) \right\} \,\pi(\beta)\nonumber\\
&= \exp\left\{ \sum_{t=1}^T y_t\cdot\theta(x_t) 
- \sum_{t=1}^T \psi(\theta(x_t)) \right\} \,\pi(\beta) \,,\nonumber
\end{align}
which rarely is available in closed form. In addition, in some cases $\psi$ may
be costly simply to compute and in others $T$ may be large or even very large. 

A standard testing situation is to decide whether or not a factor, $x_1$ say,
is influential on the dependent variable $y$. This is often translated as testing whether or not
the corresponding component of $\beta$, $\beta_1$, is {\em equal} to $0$, i.e.~$\beta_1=0$. If we denote by $\beta_{-1}$ the {\em other} components of $\beta$, the Bayes factor
for this hypothesis will be
\begin{align}
\int_{\mathbb{R}^p} \exp&\left\{ \sum_{t=1}^T y_t\cdot g(\beta^\text{T} x_t) 
- \sum_{t=1}^T \psi(g(\beta^\text{T} x_t)) \right\} \,\pi(\beta) \,\text{d}\beta\bigg/\nonumber\\
\int_{\mathbb{R}^{p-1}} &\exp\left\{ \sum_{t=1}^T y_t\cdot g(\beta^\text{T}_{-1}(x_t)_{-1}) 
-\sum_{t=1}^T \psi( \beta^\text{T}_{-1}(x_t)_{-1} ) \right\} \,\pi_{-1}(\beta_{-1} ) 
\,\text{d}\beta_{-1} \,,\nonumber
\end{align}
when $\pi_{-1}$ is the prior constructed for the null hypothesis. Obviously, except for the normal conjugate
case, both integrals cannot be computed in a closed form.
\hfill$\blacktriangleleft$
\end{example}

\subsection{Monte Carlo Solutions}

As noted above, the setting is ripe for a direct application of simulation techniques, as the underlying
probabilistic structure can exploited by either simulating pseudo-data---see, e.g., the bootstrap and the
maximum likelihood approaches---or parameter values---for the Bayesian approach---. The use of simulation
in statistics can be traced to the origins of the Monte Carlo method and simulation-based evaluations of the
power of testing procedures are available from the mid 1950's \cite{hammersley:handscomb:1964}.

It is thus no surprise that the standard Monte Carlo approximation to integrals
$$
\mathfrak{I} = \int h(x) f(x)\,\text{d}x \approx \dfrac{1}{T}\,\sum_{t=1}^T h(x_i)\,,\quad x_1,\ldots,x_T
\sim f(x)\,,
$$
and the importance sampling substitutes
$$
\mathfrak{I} \approx \dfrac{1}{T}\,\sum_{t=1}^T \dfrac{f(x)}{g(x)}\,h(x_i)\,,\quad x_1,\ldots,x_T \sim g(x)\,,
$$
are thus in use in all settings where the density $f(\cdot)$ can be easily simulated, or where a good enough
substitute density $g(\cdot)$ can instead be simulated
\cite{hammersley:handscomb:1964,rubinstein:1981,ripley:1987}.  We recall that, because $\mathfrak{I}$ can be
represented in an infinity of ways as an expectation, there is no need to simulate from the distribution with
density $f$ to get a good approximation of $\mathfrak{I}$.  For any probability density $g$ with
$\mathrm{supp}(g)$ including the support of $hf$, the integral $\mathfrak{I}$  can also be represented as an
expectation against $g$ as above.  This {\em Monte Carlo method with importance function} $g$ almost surely
converges to $\mathfrak{I}$ and the estimator is unbiased. This is classical Monte Carlo methodology
\cite{halton:1970}, however the selection of importance sampling densities $g(\cdot)$, the control of
convergence properties, the improvement of variance performances were particularly studied in the 1980's
\cite{smith:1984,geweke:1988}.

More advanced simulation techniques were however necessary to deal with large dimensions, multimodal targets,
intractable likelihoods, or/and missing data issues. Those were mostly introduced in the 1980's and 1990's,
even though precursors can be found in earlier years \cite{robert:casella:2010}, and they unsurprisingly stem
from the fringe of statistics (image analysis, signal processing, point processes, econometrics, surveys, etc.)
where the need for more powerful computational tools was more urgent. The most important advance is undoubtedly
the introduction of Markov chain Monte Carlo (MCMC) methods into statistics as it impacted the whole practice and
perception of Bayesian statistics \cite{robert:casella:2010}. We describe those methods in the next section.
More recently, a new branch of computational methods called ABC (for approximate Bayesian computations) was
launched by population geneticists to overcome computational stopping blocks due to intractable likelihoods, as
described in Section \ref{sec:ABC}.

\section{MCMC ALGORITHMS}\label{sec:MCMC}

\subsection{Markov Chain Monte Carlo methods}
As old as the Monte Carlo method itself (see \citeNP{robert:casella:2010}), the MCMC
extensions try to overcome the limitation of regular Monte Carlo methods (primarily, complexity or dimension of
the target) by simulating a Markov chain whose stationary and limiting distribution is the target distribution.
There exist rather generic ways of producing such chains, including Metropolis--Hastings and Gibbs algorithms.
Besides the fact that stationarity of the target distribution is enough to justify a simulation method by
Markov chain generation, the idea at the core of MCMC algorithms is that local exploration, when properly
weighted, can lead to a valid representation of the distribution of interest, as for instance, the
Metropolis--Hastings algorithm \cite{metropolis:rosenbluth:rosenbluth:teller:teller:1953,hastings:1970}.

Given an almost arbitrary transition distribution with density $q(x,x^\prime)$, the Markov chain is associated
with the following transition. At time $t$, given the current value $x_t$, the corresponding Markov chain explores the
surface of the target density in the neighbourhood of $x_t$ by proposing $x^*_{t+1}\sim q(x_t,\cdot)$ and then
accepting this new value with probability
$$
\dfrac{f(x^*_{t+1})\,q(x^*_{t+1},x_t)}{f(x_{t})\,q(x_t,x^*_{t+1})} \wedge 1 \,.
$$
The validation of this simple strategy is given by the detailed balance property of the Metropolis--Hastings
transition density, $\mu(x,x^\prime)$, namely that $f(x)\mu(x,x^\prime) = f(x^\prime) \mu(x^\prime,x)$.

\begin{figure}
\centerline{\includegraphics[width=9truecm]{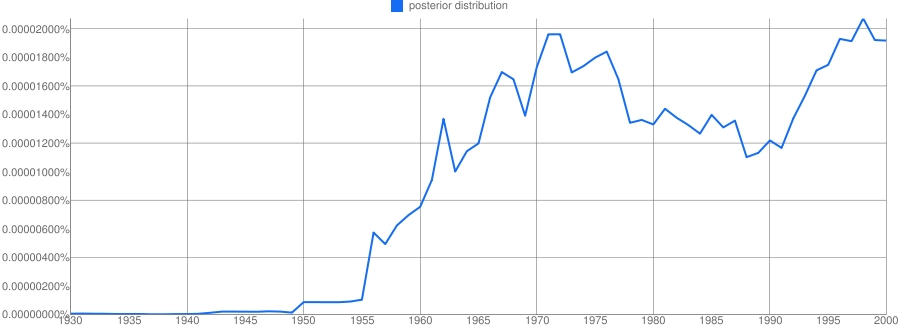}}
\caption{\label{fig:googlepost}
Occurrences of ``posterior distribution" in Google English corpus using Ngram viewer.}
\end{figure}

This MCMC simulation technique is obviously generic and thus applies in many other fields than statistics, but the
reassessment of the method in the 1980's by several statisticians (including Julian Besag, Don and Stuart
Geman, Brian Ripley, David Spiegelhalter, Martin Tanner, Alan Gelfand and Adrian Smith, as detailed in
\citeNP{robert:casella:2010}) brought a considerable impetus to the development of Bayesian statistical
methodology by providing a general way of handling high dimensional problems and complex
densities. An illustration of this surge is produced in Figure \ref{fig:googlepost} which shows how the use of the term
``posterior distribution" considerably increased after 1990, following \citeN{gelfand:smith:1990}
who advertised the Gibbs sampler as a mean of exploring posterior distributions.

The difficulty inherent to Metropolis--Hastings algorithms like the random walk version, where $q(x,x^\prime)$
is symmetric, is the scaling of the proposal distribution: the scale of $q$ must correspond to the shape of the target
distribution so that, in a reasonable number of steps,  the whole support of this distribution can be visited.
If the scale of the proposal is too small, this will not happen as the algorithm stays ``too local" and, if
there are several modes on the target, the algorithm may get trapped within one modal region because it
cannot reach other modal regions relying on jumps of too small a magnitude. The larger the dimension $p$ of the target
is, the harder it is to set up an efficient proposal, because
\begin{enumerate}\renewcommand{\theenumi}{(\alph{enumi})}
\item the curse of dimension implies that there is a larger portion of the space with zero probability;
\item the knowledge and intuition about the modal regions get weaker;
\item the scaling parameter is a symmetric $(p,p)$ matrix $\Xi$ in the
proposal $q(x,x^\prime) = g((x-x^\prime)^\text{T} \Xi (x-x^\prime))$.
Even when the matrix $\Xi$ is diagonal, it gets harder to chose as the
dimension increases.
\end{enumerate} 
Unfortunately, an on-line scaling of the algorithm by looking for instance at the empirical acceptance rate is
theoretically flawed in that it cancels the Markov validation. Furthermore, the attraction of a modal region
may give a false sense of convergence and lead to a choice of too small a scale, simply because other modes
will not be visited during the scaling experiment.

\subsection{The Challenge of Adaptivity}
Thus, given the range of situations where MCMC applies, it is unrealistic to hope for a {\em generic} MCMC
sampler that would function in every possible setting.  The reason for this ``impossibility theorem" is that,
in genuine problems, the complexity of the distribution to simulate is the very reason why MCMC is used! So
it is unrealistic to ask for a prior opinion about this distribution, its support, or the parameters of the
proposal distribution used in the MCMC algorithm: intuition is close to null in most of these problems.

However, the performances of off-the-shelf algorithms like the random-walk Metropolis--Hastings scheme bring
information about the distribution of interest or at least about the adequacy of the current proposal and, as
such, could be incorporated in the design of more efficient algorithms. The difficulty is that one usually miss
the time to train the algorithm on these previous performances.  While it is natural to think that the
information brought by the first steps of an MCMC algorithm should be used in later steps, the validation of
such a learning mechanism is complex, because of its non-Markovian nature. Usual ergodic theorems do not apply in such cases.
Further, it may be that, in practice, such algorithms do degenerate to point masses due to too a rapid
decrease in the variability of their proposal.

\begin{figure}[hbtp]
\centerline{\includegraphics[width=11cm,height=4cm]{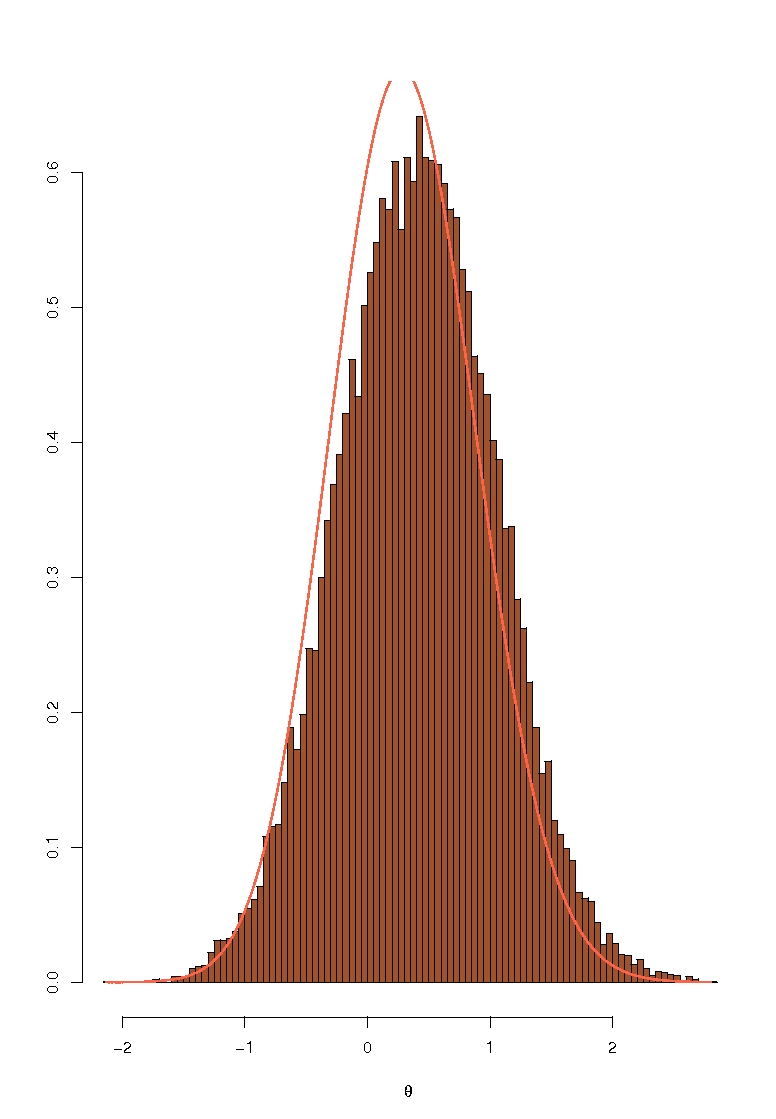}}
\caption[Invalid adaptive MCMC with large variance]{\label{fig:badap2}
Comparison of the distribution of an adaptive scheme sample 
of $25,000$ points with initial variance of $2.5$
and of the target distribution.}
\end{figure}

\begin{example}\label{ex:adatoy}
Consider a $t$-distribution $\mathcal{T}(\nu,\theta,1)$
sample $(x_1,\ldots,x_n)$ with $\nu$ known. Assume in addition a flat prior
$\pi(\theta)=1$. While the posterior distribution
on $\theta$ can be easily plotted, up to a normalising constant,
direct simulation and computation from this posterior is impossible.
In an MCMC framework, we could fit a normal proposal from the
empirical mean and variance of the previous values of the chain,
$$
\mu_t = \frac{1}{t}\,\sum_{i=1}^t \theta^{(i)}
\quad\hbox{and}\quad
\sigma^2_t = \frac{1}{t}\,\sum_{i=1}^t (\theta^{(i)} - \mu_t)^2\,.
$$
This leads to a Metropolis--Hastings algorithm with acceptance probability
$$
\prod_{j=2}^n \left[ \frac{\nu + (x_j-\theta^{(t)})^2}{\nu + (x_j-\xi)^2}
\right]^{-(\nu+1)/2} \, \frac{\exp -(\mu_t-\theta^{(t)})^2/2\sigma^2_t}{
\exp -(\mu_t-\xi)^2/2\sigma^2_t}\,,
$$
where $\xi$ is generated from $\mathcal{N}(\mu_t,\sigma^2_t)$.
The invalidity of this scheme (related to the dependence on the whole past values
of $\theta^{(i)}$) is illustrated by Figure \ref{fig:badap2}:
for an initial variance of $2.5$, there
is a bias in the fit, even after stabilisation of the empirical mean and variance.
\hfill$\blacktriangleleft$
\end{example}

The overall message is thus that one should not {\em constantly} adapt the proposal distribution on the past
performances of the simulated chain. Either the adaptation must cease after a period of {\em burnin} (not to be
taken into account for the computations of expectations and quantities related to the target distribution), or
the adaptive scheme must be theoretically assessed on its own right. If we remain within the MCMC approach
(rather than adopting sequential importance methods as in \citeNP{cappe:douc:guillin:marin:robert:2007}),
this later path is not easy and only a small
community (gathering at the {\em Adap'ski} workshops, run every three year since 2005) is working on
establishing valid adaptive schemes. Earlier works include those of \citeN{Gilks:Roberts:Sahu:1998} who use
regeneration to create block independence and preserve Markovianity on the paths rather than on the values,
\citeN{Haario:Sacksman:Tamminen:1999,Haario:Sacksman:Tamminen:2001} who derive a proper adaptation scheme in
the spirit of stochastic optimisation by using a ridge-like correction to the empirical variance, and
\citeN{Andrieu:Robert:2001} who propose a more general framework of valid adaptivity based on stochastic
optimisation and the Robbin-Monro algorithm. The latter actually embeds the chain of interest $\theta^{(t)}$ in
a larger chain $(\theta^{(t)},\xi^{(t)},\partial^{(t)})$ that also includes the parameter of the proposal
distribution as well as the gradient of a performance criterion. More recent works building on this principle
and deriving sufficient conditions for ergodicity include
\citeN{andrieu:moulines:2006,roberts:rosenthal:2007,craiu:rosenthal:yang:2009,saksman:vihola:2010,atchade:2011},
\citeN{atchade:fort:2010,ji:schmidler:2009}. They mostly contain development of a theoretical nature.

This line of research has led \citeN{roberts:rosenthal:2006} to establish guiding rules as to when an adaptive
MCMC algorithm is converging to the correct target distribution, to the point of constructing an R package
called \verb+amcmc+ \cite{rosenthal:2007}. More precisely, \citeN{roberts:rosenthal:2006} propose a {\em
diminishing adaptation} condition that states that the total variation distance between two consecutive kernels
must uniformly decrease to zero (which does not mean that the kernel must converge!). For instance, a random
walk proposal that relies on the empirical variance of the sample will satisfy this condition.  The scale of
the random walk is then tuned in each direction toward an optimal acceptance rate of $0.44$.  To this effect,
for each component of the simulated vector, a factor $\delta_i$ corresponding to the logarithm of the random
walk standard deviation is updated every $50$ iterations by adding or subtracting a factor $\epsilon_t$
depending on whether or not the average acceptance rate on that batch of $50$ iterations and for this component
was above or below $0.44$. If $\epsilon_t$ decreases to zero as $\min(.01,1/\sqrt{t})$, the conditions for
convergence are satisfied. Another package called \verb+Grapham+ was also recently developed by
\citeN{vihola:2010}.


%
%
%
%
\section{ABC METHODS}\label{sec:ABC}

\subsection{Intractable Likelihoods}
\newcommand\bx{\mathbf{x}}
\newcommand\by{\mathbf{y}}
\newcommand\bz{\mathbf{z}}
\newcommand\bu{\mathbf{u}}
When facing settings where the likelihood $\ell(\theta|\by)$ is not available in closed form, the
above solutions are not directly available. In the particular set-up of hierarchical models with partly
conjugate priors, it may be that the corresponding conditional distributions can be simulated and this property
leads to a Gibbs sampler \citeN{gelfand:smith:1990}, but such a decomposition is not available in general. In
the specific setting of latent variable models, the likelihood may be expressed as an intractable multidimensional integral
$$ 
\ell(\theta|\by) = \int \ell^\star(\theta|\by,\bu) \text{d}\bu\,,
$$
where $\by$ is observed and $\bu\in\mathbb{R}^p$ is not, while the joint distribution
$\pi(\theta,\bz|\by)\propto\pi(\theta)\ell^\star(\theta|\by,\bu)$ can be simulated.  The increase in dimension
induced by the passage from $\theta$ to $(\theta,\bu)$ may be such that the convergence properties
of the corresponding MCMC algorithms are too poor for the algorithm to be considered.

Bayesian inference thus needs to handle a large class of settings where the likelihood function is not
completely known, and where exact (or even MCMC) simulation from the corresponding posterior distribution is
impractical or even impossible. The ABC methodology, where ABC stands for {\em approximate Bayesian
computation}, offers an almost automated resolution of the difficulty with intractable-yet-simulable models.
It was first proposed in population genetics by \citeN{tavare:balding:griffith:donnelly:1997} who bypassed the
computation of the likelihood function using simulation. 
\citeN{pritchard:seielstad:perez:feldman:1999} then produced a generalisation based on an approximation of the
target.

\subsection{Approximative Resolution}
The principle of the ABC algorithm \cite{tavare:balding:griffith:donnelly:1997} is a simple accept-reject
algorithm: if one keeps simulating $\theta\sim\pi(\theta)$ and $x\sim f(x|\theta)$ until $x=y$, the resulting
$\theta$ is distributed from $\pi(\theta|y) \propto \pi(\theta)f(y|\theta)$. However, when the event $x=y$ has
probability zero, the algorithm cannot be implemented. \citeN{pritchard:seielstad:perez:feldman:1999} used
instead an approximation of the above by simulating pairs $(\theta),x$ until the condition
$$
\rho\{\eta(\bz),\eta(\by)\}\leq \epsilon
$$
is met, where  $\eta$ is a (maybe insufficient) statistic,
$\rho$ is a distance, and $\epsilon>0$ is a tolerance level.

The above algorithm is ``likelihood-free" in that it only requires the ability to simulate from the
distribution $f(x|\theta)$ and it is approximative in that it produces simulations from 
\begin{equation}\label{eq:abctarget}
\pi_\epsilon(\theta|\by)= \int_\mathcal{Z}
{\pi(\theta)f(\bz|\theta)\mathbb{I}_{A_{\epsilon,\by}}(\bz)\,\text{d}\bz}
\Big/{\int_{A_{\epsilon,\by}\times\mathbb{R}^d}\pi(\theta)f(\bz|\theta)\text{d}\bz\text{d}\theta}\,,
\end{equation}
where $\mathbb{I}_B(\cdot)$ denotes the indicator function of the set $B$ and where
$$
A_{\epsilon,\by}=\{\bz\in\mathcal{D}|\rho\{\eta(\bz),\eta(\by)\}\leq \epsilon\} \,.
$$
The basic idea behind ABC is that using a representative (enough) summary statistic $\eta$
coupled with a small (enough) tolerance $\epsilon$ should produce a good (enough) approximation to the
posterior distribution. At a deeper level, the algorithm implements a non-parametric approximation of the
conditional distribution using binary kernels \cite{beaumont:zhang:balding:2002,blum:2010,blum:francois:2010}.

\begin{figure}
\centerline{\includegraphics[width=9truecm]{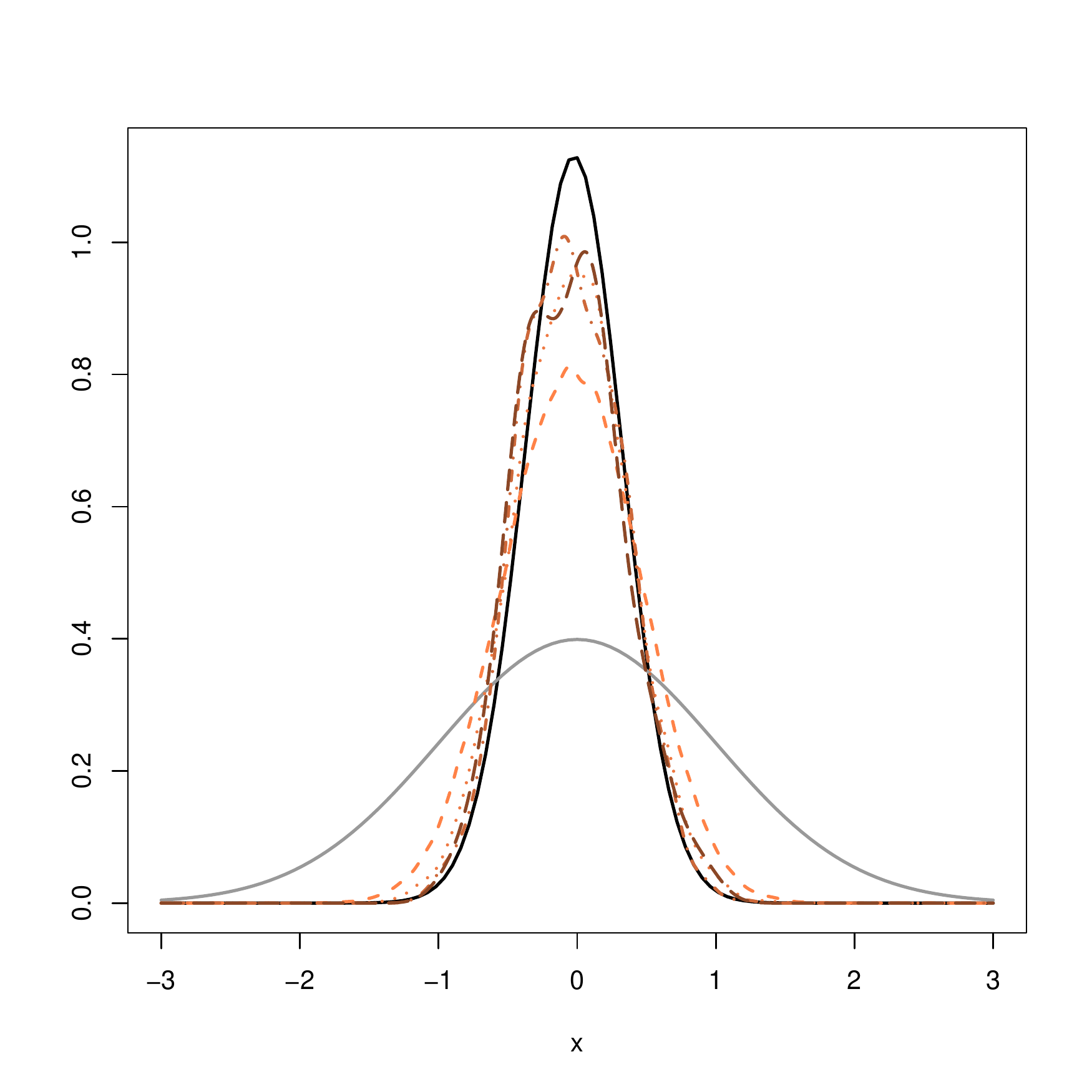}}
\caption{\label{fig:adada}
Approximations of the genuine posterior distribution {\em (in dark)} by ABC versions for $\epsilon$ equal to the $.01,.1,1,10$
\%~quantiles of $10^6$ sampled distances. {\em The prior distribution is the grey curve.}}
\end{figure}

\begin{example}\label{ex:abctoy} Considering the toy problem of Example \ref{ex:adatoy}, a
simple implementation of the ABC algorithm consists in simulating the location parameter $\theta$ from the
prior---chosen to be a $\mathcal{N}(0,1)$--- and in simulating $n$ realisations from a 
$\mathcal{T}(\nu,\theta,1)$ distribution until those $n$ realisations are close enough from the original observations.
In generic ABC algorithms, the tolerance $\epsilon$ is chosen as a quantile of the
distances between the true data and the simulated pseudo-data. Figure \ref{fig:adada} studies the impact of
this choice starting from the $10\%$ quantile down to the $.01\%$ quantile. The ABC approximations get closer
to the true posterior as $\epsilon$ decreases, the final approximation differing more for variance than for
bias reasons.
\hfill$\blacktriangleleft$
\end{example}

In more realistic problems, simulating from the prior distribution is very inefficient because it does not
account for the data at the proposal stage and thus leads to proposed values located in low posterior
probability regions. As an answer to this problem, \citeN{marjoram:etal:2003} have introduced an MCMC-ABC
algorithm targeting \eqref{eq:abctarget}. Sequential alternatives can also enhance the efficiency of the ABC
algorithm, by learning about the target distribution, as in the ABC-PMC algorithm---based on genuine importance
sampling arguments---of \citeN{beaumont:cornuet:marin:robert:2009} and the ABC-SMC algorithm---deriving a
sequential Monte Carlo (SMC) filter---of \citeN{delmoral:doucet:jasra:2009} and \citeN{drovandi:pettitt:2010}.

\subsection{Convergence and Limitations}
While the original argument of letting $\epsilon$ go to zero is clearly illustrated by the above example, more
advanced arguments have been recently produced about the convergence of ABC algorithms. Besides the
non-parametric parallels drawn in \citeN{beaumont:zhang:balding:2002,blum:2010,blum:francois:2010}, a different
perspective is found in the pseudo-likelihoods argument of
\citeN{fearnhead:prangle:2010} and \citeN{dean:singh:jasra:peters:2011}. In this perspective, ABC is an exact algorithm for
an approximate distribution, which is converging to the exact posterior as the sample size grows to infinity.

ABC being able to produce samples from posterior distributions, it does not come as a surprise that it is used
for model choice because the latter often involves a computational complexification. Estimating the posterior
probabilities of models under competition by ABC is thus proposed in most of the literature
(see, e.g.,
\citeNP{cornuet:santos:beaumont:etal:2008,grelaud:marin:robert:rodolphe:tally:2009,toni:etal:2009,toni:stumpf:2010}).
However, it has been recently exposed in
\citeN{didelot:everitt:johansen:lawson:2011,robert:cornuet:marin:pillai:2011} that those approximations may
fail to converge in the case the summary statistics are not sufficient for model comparison. More empirical
evaluations as those tested in \citeN{ratmann:andrieu:wiujf:richardson:2009} should thus be implemented to
assess the relevance of those approximations, unless one uses the whole data instead of summary statistics.

\section{CONCLUSION}
The present tutorial is naturally both very incomplete and quite partial. Even within Bayesian methods, one
glaring omission is the area of sequential Monte Carlo methods (or particle filters) used to analyse non-linear
non-Gaussian state space models. While particle filters and improvements have been around for several years
\cite{gordon:salmon:smith:1993,pitt:shephard:1999,delmoral:doucet:jasra:2006}, handling unknown parameters as
well is a much harder goal and it is only recently that significant progress has been made with the particle
MCMC method of \citeN{andrieu:doucet:holenstein:2010}. This new class of MCMC algorithms relies on a population
of particles produced by an SMC algorithm to build an efficient proposal at each MCMC iteration. (That this
approximation remains a valid MCMC algorithm is quite an involved resul, in connection with the results of
\citeNP{andrieu:roberts:2009}). This area is
currently quite active with alternatives and softwares being developed. More generally, parallel computing is
now used in a growing fraction of applications, particularly in genetics, with a direct impact on the nature of
the simulation methods \cite{tom:sinsheimer:suchard:2010,jacob:robert:smith:2010}, which include more and more
non-parametric aspects \cite{jordan:2010,hjort:holmes:mueller:walker:2010}.

Simulation techniques are clearly found in too many of current statistical methods to be even mentioned here
and the trend is definitely upward. \citeN{speed:2011} concludes ``we are heading to an era when all
statistical analysis can be done by simulation". This is indeed quite an exciting time for computational
statisticians!

\section*{ACKNOWLEDGEMENTS}
This research is supported partly by the Agence Nationale de la Recherche through the 2009-2012 grants {\sf Big
MC} and {\sf EMILE} and partly by the Institut Universitaire de France (IUF).


\section*{AUTHOR BIOGRAPHY}
 
\noindent {\bf CHRISTIAN P.~ROBERT} is Professor in the Department of Applied Mathematics at the Universit{\'e}
Paris-Dauphine since 2000. He is also a 2010-2015 senior member of the Institut Universitaire de France, and
the former Head of the Statistics Laboratory of the Centre de Recherche en Economie et Statistique (CREST). He
was previously Professor at the Universit{\'e} de Rouen from 1992 till 2000 and has held visiting positions in
Purdue University, Cornell University, and the University of Canterbury, Christchurch, New-Zealand.  He 
is currently an adjunct professor in the Department of Mathematics and Statistics at the Queensland University
of Technology (QUT), Brisbane, Australia. He is a Fellow of the Royal Statistical Society and of the Institute
of Mathematical Statistics (IMS), as well as a Medallion Lecturer of the IMS. He was Editor of the Journal of
the Royal Statistical Society Series B from 2006 till 2009 and has been an associate editor of the {\em Annals
of Statistics}, {\em Journal of the American Statistical Society}, {\em Statistical Science}, {\em Sankhya}. He
is currently an Area Editor for the {\em ACM Transactions on Modeling and Computer Simulation} (TOMACS)
journal. He was the 2008 President of the International Society for Bayesian Analysis (ISBA). His research
areas cover Bayesian statistics, with a focus on decision theory and model selection, numerical probability,
with works cantering on the application of Markov chain theory to simulation, and computational statistics,
developing and evaluating new methodologies for the analysis of statistical models.  

\end{document}